\documentclass[conference,letterpaper]{IEEEtran}

\usepackage[utf8]{inputenc} 
\usepackage[T1]{fontenc}
\usepackage{url}
\usepackage{ifthen}
\usepackage{cite}
\usepackage[cmex10]{amsmath} 

\newtheorem{teo}{Theorem}
\newtheorem{prop}{Proposition}
\newtheorem{con}{Conjecture}

\newtheorem{rem}{Remark}

\newtheorem{lemma}{Lemma}

\DeclareFontFamily{U}{mathx}{\hyphenchar\font45}
\DeclareFontShape{U}{mathx}{m}{n}{
      <5> <6> <7> <8> <9> <10>
      <10.95> <12> <14.4> <17.28> <20.74> <24.88>
      mathx10
      }{}
\DeclareSymbolFont{mathx}{U}{mathx}{m}{n}
\DeclareFontSubstitution{U}{mathx}{m}{n}
\DeclareMathAccent{\widecheck}{0}{mathx}{"71}
\DeclareMathAccent{\wideparen}{0}{mathx}{"75}

\usepackage{tabularx,booktabs}
\usepackage[pdftex]{graphicx}
\usepackage{amsmath}
\usepackage[cmintegrals]{newtxmath}

\begin{document}
%
\title{New upper bounds for $(b,k)$-hashing}
%
%
%


 \author{%
   \IEEEauthorblockN{Stefano Della Fiore,
                     Simone Costa,
                     Marco Dalai,~\IEEEmembership{Senior~Member,~IEEE}}
                     
   \IEEEauthorblockA{                        Department of Information Engineering, 
                     University of Brescia\\
\{s.dellafiore001, simone.costa, marco.dalai\}@unibs.it}
 }

\maketitle
\begin{abstract}
For fixed integers $b\geq k$, the problem of perfect $(b,k)$-hashing asks for the asymptotic growth of largest subsets of $\{1,2,\ldots,b\}^n$ such that for any $k$ distinct elements in the set, there is a coordinate where they all differ. 

An important asymptotic upper bound for general $b, k$, was derived by Fredman and Koml\'os in the '80s and  improved for certain $b\neq k$ by K\"orner and Marton and by Arikan. 
Only very recently better bounds were derived for the general $b,k$ case by Guruswami and Riazanov, while stronger results for small values of $b=k$ were obtained by Arikan, by Dalai, Guruswami and Radhakrishnan and by Costa and Dalai.

In this paper, we both show how some of the latter results extend to $b\neq k$ and further strengthen the bounds for some specific small values of $b$ and $k$. The method we use, which depends on the reduction of an optimization problem to a finite number of cases, shows that further results might be obtained by refined arguments at the expense of higher complexity.
\end{abstract}

\begin{IEEEkeywords}
perfect hashing, list decoding, zero-error capacity
\end{IEEEkeywords}

%
\IEEEpeerreviewmaketitle

\ifCLASSOPTIONcaptionsoff
  \newpage
\fi

\section{Introduction}

\begin{figure}[b]
\centering
\includegraphics[scale=0.8]{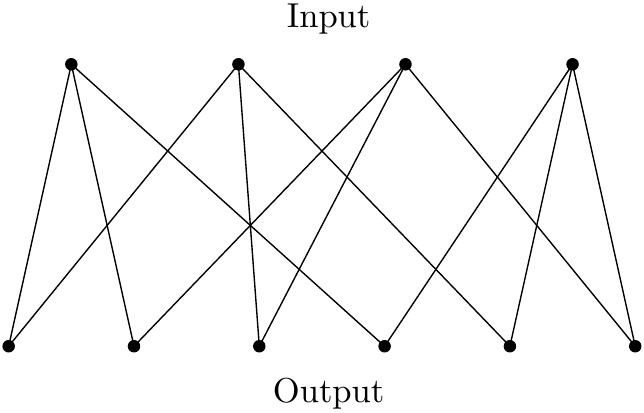}
\caption{A $4/2$ channel. Edges represent positive probabilities. Here, zero-error communication is possible  when decoding with list-size equal to $2$.}
\label{fig:channel}
\end{figure}
Let $b$, $k$ and $n$ be integers, with $b \geq k$, and let $\mathcal{C}$ be a subset of $\{1,2,\ldots,b\}^n$ with the property that for any $k$ distinct elements we can find a coordinate where they all differ. Such a set can be interpreted, by looking at it coordinate-wise, as a family of $n$ hashing functions on some universe of size $|\mathcal{C}|$. The required property then says that the family is a perfect hash family, that is, any $k$ elements in the universe are $k$-partitioned by at least one function. Alternatively $\mathcal{C}$ can be interpreted as a code of rate $\frac{1}{n}\log|\mathcal{C}|$ for communication over a channel with $b$ inputs. 
Assume that the channels is a $b/(k-1)$ channel, meaning that any $k-1$ of the $b$ inputs share one output but no $k$ distinct inputs do (see Figure \ref{fig:channel}). The required property for $\mathcal{C}$ is what is needed for the code to be a zero-error code when list decoding with list-size $k-1$ is allowed.  We refer the reader to \cite{Elias1}, \cite{FredmanKomlos}, \cite{Korner3}, \cite{nilli} and \cite{Jaikumar2} for an overview of the the more general context of this problem.

We will call any subset $\mathcal{C}$ of $\{1,2,\ldots,b\}^n$ with the described property a $(b,k)$-hash code. For the reasons mentioned above, bounding the size of $(b,k)$-hash codes is a combinatorial problem which has been of interest both in computer science and information theory. It is known that $(b,k)$-hash codes of exponential size in $n$ can be constructed and the quantity of interest is usually the rate of such codes. We will thus study the quantity
\begin{equation}
R_{(b,k)}=\limsup_{n\to \infty}\frac{1}{n}\log|\mathcal{C}_n|\,,
\end{equation}
where the $\mathcal{C}_n$ are $(b,k)$-hash codes of length $n$ with maximal rate. Note that, throughout, all logarithms are to base 2. Few lower bounds on $R_{(b,k)}$ are known. First results in this sense were given by \cite{FredmanKomlos}, \cite{Elias1} and a better bound was derived in \cite{Korner2} for $(b, k)=(3,3)$. More recently, new lower bounds were derived in \cite{xing-yuang} for infinitely many other values of $k$. 
The first, landmark result concerning upper bounds was obtained by Fredman and Koml\'os \cite{FredmanKomlos}, who showed that 
\begin{equation}
R_{(b,k)} \leq  \frac{b^{\underline{k-1}}}{b^{k-1}} \log (b-k+2)\,,
\label{eq:fredmankomlos}
\end{equation}
where $b^{\underline{k-1}} = b (b-1) \cdots (b-k+2)$.
Progresses have since been rare. A generalization of the bound given in equation \eqref{eq:fredmankomlos} was derived by K\"orner and Marton \cite{Korner2} in the form
\begin{equation}
R_{(b,k)} \leq \min_{2 \leq j \leq k-2} \frac{b^{\underline{j+1}}}{b^{j+1}} \log \frac{b-j}{k-j-1}\,.
\label{eq:kornermarton}
\end{equation}
This was further improved for different values of $b$ and $k$ by Arikan \cite{Arikan}.
In the case $b=k$, an improvement was first obtained for $k=4$ in \cite{Arikan2} and then in \cite{DalaiVenkatJaikumar}, \cite{DalaiVenkatJaikumar2}. It was proved only recently in \cite{venkat} that the Fredman-Koml\'os bound is not tight for any $k>3$; explicit better values were given there for $k=5,6$, and for larger $k$ modulo a conjecture which is proved in \cite{costaDalai}, where further improvements are also obtained for $k=5, 6$. 

In this paper, we develop a new strategy to attack some of the cases which appear not to be optimally handled by those methods, obtaining new bounds for $b=k=5,\ldots, 8$. Furthermore, we also show that our procedure improves on the existing literature for some $b\neq k$ cases, among which for example $(b, k)=(6,5)$, $(9,8)$, $(10,9)$, $(11,10)$. In order to evaluate in a fair way these $b\neq k$ cases, we first analyze the results (not derived in the referenced papers) which are obtained when the methods of 
\cite{DalaiVenkatJaikumar} and \cite{costaDalai} are extended to $b\neq k$, and compare them with the ones of \cite{Korner2}, \cite{Arikan} and \cite{venkat}. 

The generalization of the procedure used in \cite{DalaiVenkatJaikumar} is rather easy\footnote{The interested reader will find, upon inspection of the proof of Theorem 3 in \cite{DalaiVenkatJaikumar}, that modulo using a hypergraph version of the Hansel Lemma, the only new condition to check is that the upper bound given in \eqref{eq:dalaivenkatbound} is greater than $\log \frac{2b-2}{2b-3}$ for every $b \geq k \geq 4$.} and it provides us the following bound
\begin{equation}
	R_{(b,k)}  \leq \left(\frac{1}{\log b} + \frac{b^2}{(b^2-3b+2) \log \frac{b-2}{k-3}}  \right)^{-1}.
	\label{eq:dalaivenkatbound}
\end{equation}
In Table \ref{tab:bkbounds} we give a comparison between the bounds \eqref{eq:dalaivenkatbound} and \eqref{eq:kornermarton}, the bounds from \cite{Arikan} and \cite{venkat} and the generalized bound from \cite{costaDalai} for different values of $b$ and $k$. The integers in the parentheses for the bound \eqref{eq:kornermarton} represent the minimizing $j$; a parameter $j$ with the same role is involved in the other bounds and it will be discussed later. For the bounds of \cite{costaDalai}, \cite{Arikan} and \cite{venkat} it is equal to $k-2$, while for the bound of \cite{DalaiVenkatJaikumar} it is equal to $2$.
 
In Table \ref{tab:newbkbounds} we compare our new bounds with the best known bounds for $b=k=5,\ldots,8$ and for $(b,k) = (6,5)$, $(9,8)$, $(10,9)$, $(11, 10)$.

\begin{table}[ht!]
\def\arraystretch{1.17}
\caption{Upper bounds on $R_{(b,k)}$. All numbers are rounded upwards.}
\centering
\begin{tabularx}{\linewidth}{c@{\extracolsep{\fill}}c@{\extracolsep{\fill}}c@{\extracolsep{\fill}}c@{\extracolsep{\fill}}c@{\extracolsep{\fill}}c}
$(b,k)$ & \cite{costaDalai}* & \cite{DalaiVenkatJaikumar}* & \cite{Arikan} & \cite{venkat} & \cite{Korner2} \\
\hline
$(5,4)$ & 0.66126 &  \textbf{0.57303} & 0.61142 & 0.74834  & 0.73697(0)\\
$(6,4)$ & 0.87963 & \textbf{0.77709} & 0.83904 & 1.09604 & 1.00000(0)\\
$(7,4)$ & 1.03711 & \textbf{0.94372}  & 1.02931 & 1.40593  &  1.22239(0) \\
$(5,5)$ & \textbf{0.16964} & 0.25050 & 0.23560 & 0.19079 & 0.19200(3)\\
$(6,5)$ & \textbf{0.34597} & 0.45728 & 0.44149 & 0.43207 &  0.44027(3)\\
$(6,6)$ & \textbf{0.08760} & 0.21170 & 0.15484 & 0.09228 & 0.09260(4)\\
$(7,6)$ & \textbf{0.19897} & 0.38873 & 0.30554 & 0.23524 & 0.23765(4) \\
$(8,6)$ & \textbf{0.31799} & 0.53847  & 0.44888 & 0.40330 & 0.41016(4) \\
$(7,7)$ & \textbf{0.04379} & 0.18417 & 0.09747 & 0.04279 & 0.04284(5)\\
$(8,7)$ & \textbf{0.10865} & 0.34034 & 0.20340 & 0.12134 & 0.12189(5) \\
$(9,7)$ & \textbf{0.19054} & 0.47461 & 0.31204 & 0.22547 & 0.22761(5) \\
$(8,8)$ & \textbf{0.02077} & 0.16323 & 0.05769 & 0.01922 & 0.01923(6)\\
$(9,8)$ & \textbf{0.05686} & 0.30348 & 0.12874 & 0.06001 & 0.06013(6) \\
$(10,8)$ & \textbf{0.10791} & 0.42566 & 0.20754 & 0.12048 & 0.12096(6) \\
$(10,9)$ & 0.02889 & 0.27417 & 0.07668 & \textbf{0.02874} & 0.02876(7) \\
$(11,10)$ & 0.01407 & 0.25018 & 0.04289 & \textbf{0.01342} & 0.01343(8) \\
\hline
\vspace{-2mm}
\end{tabularx}
\label{tab:bkbounds}
\raggedright
\footnotesize{$*$ The generalized bound for the $(b,k)$ case}
\end{table}


\begin{table}[ht!]
\def\arraystretch{1.15}
\caption{Upper bounds on $R_{(b,k)}$. All numbers are rounded upwards.}
\centering
\begin{tabularx}{\linewidth}{c@{\extracolsep{\fill}}l@{\extracolsep{\fill}}l@{\extracolsep{\fill}}l@{\extracolsep{\fill}}l@{\extracolsep{\fill}}l@{\extracolsep{\fill}}l}
$(b,k)$ & This work & \cite{costaDalai} & \cite{DalaiVenkatJaikumar} & \cite{Arikan} & \cite{venkat}\\
\hline
$(5,5)$ & \textbf{0.16894} & 0.16964 & 0.25050 & 0.23560 & 0.19079\\
$(6,5)$ & \textbf{0.34512} & 0.34597 & 0.45728 & 0.44149 & 0.43207\\
$(6,6)$ & \textbf{0.08475} & 0.08760 & 0.21170 & 0.15484 & 0.09228\\
$(7,7)$ & \textbf{0.04090} & 0.04379 & 0.18417 & 0.09747 & 0.04279\\
$(8,8)$ & \textbf{0.01889} & 0.02077 & 0.16323 & 0.05769 & 0.01922\\
$(9,8)$ & \textbf{0.05616} & 0.05686 & 0.30348 & 0.12874 & 0.06001 \\
$(10,9)$ & \textbf{0.02773} & 0.02889 & 0.27417 & 0.07668 & 0.02874 \\
$(11,10)$ & \textbf{0.01321} & 0.01407 & 0.25018 & 0.04289 & 0.01342 \\
\hline
\end{tabularx}
\label{tab:newbkbounds}
\end{table}

The paper is structured as follows. In the Section \ref{background} we give the general structure of the method used in the mentioned recent series of works to find upper bounds using the hypergraph version of the Hansel's lemma. In Section \ref{ClusterBound} we present the main new ingredient of this paper, which is a way to improve the bounds derived in \cite{costaDalai} by means of a more careful analysis of a quadratic form that was also objective of that study. In Section \ref{CaseReduction}, we show how this idea can be effectively implemented after an appropriate reduction of the problem to a list of cases that can be studied exhaustively.

\section{Structure of the General Method}
\label{background}
The best upper bounds on $R_{(b,k)}$ available in the literature can all be seen as different applications of a central idea, which is the study of $(b,k)$-hashing by comparison with a combinations of binary partitions. This main line of approach to the problem comes from the original work of Fredman and K\'omlos \cite{FredmanKomlos}. A clear and productive formulation of the idea was given by Radhakrishnan in terms of Hansel's lemma \cite{Jkumar}, which remained the main tool used in all recent results \cite{DalaiVenkatJaikumar2}, \cite{venkat} and \cite{costaDalai}. We state the Lemma here and briefly revise for the reader convenience how this was applied in those works.

\begin{lemma}[Hansel for Hypergraphs \cite{hansel}, \cite{nilli}]
Let $K_r^d$ be a complete $d$-uniform hypergraph on $r$ vertices and let $G_1,\ldots,G_m$ be $c$-partite $d$-uniform hypergraphs on those same vertices such that $\cup_{i}G_i=K_r^d$. Let $\tau(G_i)$ be the number of non-isolated vertices in $G_i$. Then
\begin{equation}
\log \frac{c}{d-1} \sum_{i=1}^m \tau(G_i) \geq \log \frac{r}{d-1} \,.
\end{equation}
\end{lemma}

The application to $(b,k)$-hashing relies on the following observation.
Given a $(b,k)$-hash code $C$, fix any $j$ elements $x_1,x_2,\ldots, x_{j}$ in $C$, with $j=2, \ldots, k-2$. For any coordinate $i$ let $G_i^{x_1, \ldots, x_j}$ be the $(b-j)$-partite $(k-j)$-uniform hypergraph with vertex set $G\setminus\{x_1,x_2,\ldots, x_j\}$ and edge set
\begin{align}
E = & \nonumber \big\{ (y_1, \ldots, y_{k-j}) : \\ &x_{1,i}, \ldots, x_{j,i},y_{1,i}, \ldots,y_{k-j, i} \mbox{ are all distinct} \big\}\,.\label{eq:FKgraph}
\end{align}
Since $C$ is a $(b,k)$-hash code, then $\bigcup_i G_i^{x_1, \ldots, x_{j}}$ is the complete $(k-j)$-uniform hypergraph on $G\setminus\{x_1,x_2,\ldots, x_j\}$ and so
\begin{equation}
\log \frac{b-j}{k-j-1} \sum_{i=1}^n \tau(G_i^{x_1, \ldots, x_{j}})\geq \log\frac{|C|-j}{k-j-1}\,.
\label{eq:hansel_hash}
\end{equation}
This inequality allows one to upper bound $|C|$ by upper bounding the left hand side. Inequality (\ref{eq:hansel_hash}) holds for any choice of $x_1,x_2,\ldots, x_j$, so the main goal is proving that the left hand side is not too large for all possible choices of $x_1,x_2,\ldots, x_j$. The choice can be deterministic or we can take the expectation over any random selection.

Note that if the $x_{1,i},x_{2,i},\ldots, x_{j,i}$ are not all distinct (let us say that they ``collide'') then the hypergraph in \eqref{eq:FKgraph} is empty, that is the corresponding $\tau$ in the left hand side of \eqref{eq:hansel_hash} is zero. So, using codewords $x_1,x_2,\ldots, x_j$ which collide in many coordinates helps in upper bounding $|\mathcal{C}|$. On the other hand, in a coordinate $i$ where the codewords do \emph{not} collide, $\tau(G_i^{x_1, \ldots, x_j})$ depends on what a fraction of the code uses the remaining $b-j$ symbols in the alphabet. This can be made small ``on average'' if $x_1, \ldots, x_j$ are picked randomly. More precisely, let  $f_{i}$ be probability distribution of the $i$-th coordinate of $C$, that is, $f_{i,a}$ is the fraction of elements of $C$ whose $i$-th coordinate is $a$. Then, we have 
\begin{multline}
\tau(G_i^{x_1, \ldots, x_j})=\\\hspace{0.3cm}
\begin{cases}
0 \hspace{1cm} x_1,\ldots,x_{j} \mbox{ collide in coordinate }i\\
\left(\frac{|C|}{|C|-j}\right)\left(1-\sum_{h=1}^{j}f_{i,x_{h i}}\right) \hspace{0.5cm}\mbox{otherwise}
\end{cases}.
\end{multline}

So, one can make the left hand side in \eqref{eq:hansel_hash} small by using $x_{1}, \ldots,x_j$ which collide in many coordinates and at the same time have in the remaining coordinates symbols $x_{h i}$ for which the $f_{i,x_{h i}}$ are not too small. This can be obtained ``on average'' if $x_{1}, \ldots,x_{j}$ are picked in some random way over the code, since this will force values with large $f_{i,x_{h i}}$ to a appear frequently as the $i$-th coordinate in some of the $x_{1}, \ldots,x_{j}$.
There are different ways to turn this into a precise agrument to bound the right hand side of \eqref{eq:hansel_hash}. We refer the reader to \cite{costaDalai} for a detailed discussion, and we only discuss here the procedure as used there, since it is the base for our current contribution. 

The idea is to partition the code $\mathcal{C}$ in subcodes $\mathcal{C}_\omega$, $\omega\in\Omega$. The only requirement is that each subcode has size which grows unbounded with $n$ and uses in any of its first $\ell$ coordinates only $(j-1)$ symbols.
It can be show, by an easy extension of the method used for the case $b=k$ and $j=k-2$ in  \cite{costaDalai}, that if the original code has rate $R$, then for any $\epsilon>0$ one can do this with a choice of $\ell=n(R-\epsilon)/\log\left(\frac{b}{j-1}\right)$ for $n$ large enough. Given such a partition of our code, if we select codewords $x_{1}, \ldots,x_{j}$ within the same subcode $\mathcal{C}_\omega$, they will collide in the first $\ell$ coordinates and the corresponding contribution to the l.h.s. of \eqref{eq:hansel_hash} will be zero. We then add the randomization. We pick randomly one of the subcodes $\mathcal{C}_\omega$ and randomly select the codewords $x_{1}, \ldots,x_{j}$ within $\mathcal{C}_\omega$. We then upper bound the expected value of the left hand side of \eqref{eq:hansel_hash} under this random selection to obtain an upper bound on $|\mathcal{C}|$, that is
\begin{align}
\log &\frac{|C|-j}{k-j-1} \nonumber \\ &\leq \log \frac{b-j}{k-j-1}  \mathbb{E}_{\omega}(\mathbb{E}[\sum_{i\in [\ell+1,n]}\tau(G_i^{x_1,x_2,\dots,x_{j}})|\omega])  \nonumber \\
&= \log \frac{b-j}{k-j-1}  \sum_{i\in [\ell+1,n]}\mathbb{E}_{\omega}(\mathbb{E}[\tau(G_i^{x_1,x_2,\dots,x_{j}})|\omega])\label{eq:sumellton}.
\end{align}
Here, each subcode $\mathcal{C}_\omega$ is taken with probability $\lambda_{\omega}=|\mathcal{C}_\omega|/|\mathcal{C}|$, and $x_{1}, \ldots,x_{j}$ are taken uniformly at random (without repetitions) from $\mathcal{C}_\omega$. 

As mentioned before, let $f_{i}$ be the probability distribution of the $i$-th coordinate of $C$, and let instead $f_{i|\omega}$ be the distribution of the $i$-th coordinate of the subcode $C_\omega$ (with components, say, $f_{i,a|\omega}$) . Then, for $i>\ell$, we can write
\begin{align}
\mathbb{E} &[ \tau(G_i^{x_1, \ldots, x_{j}})|\omega] =\left(1+o(1)\right) \nonumber \\ 
&\sum_{\stackrel{\text{distinct }}{ a_1,\ldots,a_{j}}} f_{i,a_1|\omega}f_{i,a_2|\omega}\cdots f_{i,a_{j}|\omega}(1-f_{i,a_1}-\cdots-f_{i,a_{j}} )\label{eq:exptaugf}
\end{align}
where the $o(1)$ is meant as $n\to\infty$ and is due, under the assumption that $C_\omega$ grows unbounded with $n$, to sampling without replacement within $C_\omega$.
Now, since $\lambda_{\omega}=|\mathcal{C}_\omega|/|\mathcal{C}|$, $f_i$ is actually the expectation of $f_{i|\omega}$ over the random $\omega$, that is, using a different dummy variable $\mu$ to index the subcodes for convenience,
$$
f_i=\sum_{\mu}\lambda_{\mu}f_{i|\mu}\,.
$$
Using this in \eqref{eq:exptaugf}, one notices that when taking further expectation over $\omega$ it is possible to operate a symmetrization in $\omega$ and $\mu$. If we denote with $\Psi$ for the polynomial function defined for two probability distribution $p=(p_1,p_2,\dots,p_b)$ and $q=(q_1,q_2,\dots,q_b)$ as
\begin{align}
\Psi(p,q) = &\frac{1}{(b-j-1)!} \\ \nonumber \sum_{\sigma\in S_b} &p_{\sigma(1)}p_{\sigma(2)}\dots p_{\sigma(j)}q_{\sigma(j+1)} + \\ &q_{\sigma(1)}q_{\sigma(2)}\dots q_{\sigma(j)}p_{\sigma(j+1)}.\label{eq:defPsi}
\end{align}
Then the expectation of \eqref{eq:exptaugf} over $\omega$ can be written as
\begin{align}
\mathbb{E}[\tau (G_i^{x_1,x_2,\dots,x_{j}})] =\left(1+o(1)\right)\frac{1}{2}\sum_{\omega,\mu\in \Omega}\lambda_{\omega}\lambda_{\mu}\Psi(f_{i|\omega},f_{i|\mu}).\label{simmetrizzata2}
\end{align}

In \cite{costaDalai}, the global maximum of the function $\Psi(p,q)$, over arbitrary distributions $p$ and $q$, say
\begin{equation}
\label{eq:PsiMax}
\Psi_{\max}=\max_{p,q}\Psi(p,q)\,,
\end{equation}
 was used to deduce the inequality, valid for any $i>\ell$,
\begin{equation}
\mathbb{E}[\tau (G_i^{x_1,x_2,\dots,x_{j}})]\leq (1+o(1))\frac{1}{2}\Psi_{\max}\,.
\end{equation}
Then
\begin{equation}
\log{|C|} \leq (1+o(1)) \frac{1}{2} (n-\ell) \Psi_{\max} \log \frac{b-j}{k-j-1} \,,
\end{equation}
from which, using the value of $\ell$ described above, one deduces
\begin{align*}
R\leq (1+o(1))\frac{1}{2}\left[1-\frac{R}{\log\left(\frac{b}{j-1}\right)}\right]\Psi_{\max} \log \frac{b-j}{k-j-1}.
\end{align*}
This gives the explicit bound
\begin{align}
R_{(b,k)} \leq \frac{1}{\frac{2}{\Psi_{\max} \log \frac{b-j}{k-j-1} }+\frac{1}{\log\left(\frac{b}{j-1}\right)}}\,.
\label{eq:RboundedbyPsi}
\end{align}

A weakness in this bound comes from the fact that distributions $p$ and $q$ that maximize $\Psi(p,q)$ could exhibit some opposing asymmetries, in the sense that they give higher probabilities to different symbols. When used as a replacement for \emph{each} of the pairs of $f_{i|\omega}$ and $f_{i|\mu}$ in \eqref{simmetrizzata2}, we have a rather conservative bound, because pairs $(p,q)$ which give high values for $\Psi(p,q)$ will give low values for $\Psi(p;p)$ and $\Psi(q;q)$, and equation \eqref{simmetrizzata2} contains a weighted contribution from all pairings of $f_{i|\omega}$ and $f_{i|\mu}$. In other words, observed that \eqref{simmetrizzata2} is a quadratic form in the distribution $\lambda$ with kernel $\Psi(p,q)$, if the kernel has maximum value $\Psi_{\max}$ in some off-diagonal $(p,q)$-positions to which there correspond small ``in-diagonal'' values at $(p,p)$ and $(q,q)$, then using  $\Psi_{\max}$ as a bound for the whole quadratic form can be quite a conservative approach.

In this paper, we approach \eqref{simmetrizzata2} more carefully by 
clustering the possible distributions $f_{i|\omega}$ in different groups depending on how balanced or unbalanced they are, and bounding $\Psi(f_{i|\omega},f_{i|\mu})$ for $f_{i|\omega}$ and $f_{i|\mu}$ in those different groups. From this, we deduce a bound on the quadratic form. Note that since in the problem under consideration (that is, as $n\to\infty$) we have no limit in the granularity of the distributions $f_{i,\omega}$, the quadratic form that we have to bound might in principle have a limiting value which is only achieved with a continuous distribution $\lambda$ over the simplex of $b$-dimensional distributions $\mathcal{P}_b$. Still, once we consider a finite number of clusters $r$ for the distributions $f_{i|\omega}$, our quadratic form is upper bounded by a corresponding $r$-dimensional one. In our derivation, we will use $b+1$ clusters with some symmetric structure which allows us to further reduce the complexity to an equivalent four dimensional form and then to a quadratics in one single variable.

\section{Bounding the quadratic form}
\label{ClusterBound}
Based on the discussion in the previous Section, we now enter the problem of determining better upper bounds on the right hand side of \eqref{simmetrizzata2}. We simplify here the notation and consider the quadratic form
\begin{equation}
\sum_{p,q}\lambda_p\lambda_q \Psi(p,q)
\label{eq:QuadraticForm}
\end{equation}
where $p$ and $q$ run over an arbitrary  finite set of points in the simplex $\mathcal{P}_b$ of $b$-dimensional probability distribution and $\lambda$ is a probability distribution over such set. We consider partitions of $\mathcal{P}_b$ in disjoint subsets to find upper bounds on the quadratic form \eqref{eq:QuadraticForm} in terms of simpler ones. If we have a partition $\{\mathcal{P}_b^0,\mathcal{P}_b^1,\ldots,\mathcal{P}_b^r\}$ of $\mathcal{P}_b$ and we define
$$
m_{i,h}=\sup_{p\in \mathcal{P}_b^i , q\in\mathcal{P}_b^h} \Psi(p,q)\,,\qquad \eta_i=\sum_{p\in\mathcal{P}_b^i}\lambda_p\,,
$$
then clearly 
\begin{align}
\sum_{p,q}\lambda_p\lambda_q \Psi(p,q) & \leq \sum_{i,h}\sum_{p\in\mathcal{P}_b^i}\sum_{q\in\mathcal{P}_b^h}\lambda_p \lambda_q m_{i,h}\nonumber\\
& \leq \sum_{i,h}\eta_i \eta_h m_{i,h}\,.\label{eq:ReducedQuadratic}
\end{align}
This is a convenient simplification since we have now an $r$-dimensional problem which we might be able to deal with in some computationally feasible way. We will use this procedure with two different partitions in terms of how balanced or unbalanced the distributions are.
We take $b+1$ subsets with some symmetry which allows us to further reduce the complexity.

\textbf{Partition based on maximum value.} We first consider a partition of $\mathcal{P}_b$ in terms of the largest probability value which appears in a distribution. We use a parameter $\epsilon<1/(b-1)$; all quantities will depend on $\epsilon$ but we do not write this in order to avoid cluttering the notation. We define $b$ sets of unbalanced distributions
$$
\widecheck{\mathcal{P}}_b^{i} = \left\{p\in \mathcal{P}_b:p_i>1-\epsilon\right\}\,
$$
for every $1\leq i \leq b$, and correspondingly a set of balanced distributions
$$
 \widecheck{\mathcal{P}}_b^{0} = \left\{p\in \mathcal{P}_b:p_i\leq 1-\epsilon\ \forall i\right\}\,.
$$
Note that these are all disjoint sets since $\epsilon<1/(b-1)$. Following the scheme mentioned above, we can consider the values $m_{i,h}$ and $\eta_i$ for this specific partition. However, due to symmetry, the values $m_{i,h}$ can be reduced to only four cases, depending on whether $p$ and $q$ are both balanced, one balanced and one unbalanced, or both unbalanced, either on the same coordinate or on different coordinates.\\
Assuming $1\leq i,h\leq b$ with $i\neq h$, the following quantities are then well defined and independent of the specific values chosen for $i$ and $h$
\begin{equation}
\begin{aligned}
\widecheck{M}_1  &=\sup_{p,q \in \widecheck{\mathcal{P}}_b^{0}} \Psi(p,q) &\qquad &
\widecheck{M}_2  &=\sup_{p \in \widecheck{\mathcal{P}}_b^{0},q\in \widecheck{\mathcal{P}}_b^{i}} \Psi(p,q)\\
\widecheck{M}_3  &=\sup_{p,q \in \widecheck{\mathcal{P}}_b^{i}} \Psi(p,q) &\qquad &
\widecheck{M}_4  &=\sup_{p \in \widecheck{\mathcal{P}}_b^{i},q\in \widecheck{\mathcal{P}}_b^{h}} \Psi(p,q)
\end{aligned}
\label{eq:MaxMis}
\end{equation}
These values can then be used in \eqref{eq:ReducedQuadratic} in place of the values $m_{i,h}$.

\textbf{Partition based on the minimum value.} We also consider a partition of $\mathcal{P}_b$ using constraints from below. Again we use a parameter $\epsilon$ which will be then tuned. We assume here $\epsilon<1/b$. Consider now the following disjoint sets of unbalanced distributions
$$
\widehat{\mathcal{P}}_b^{i} = \left\{p\in \mathcal{P}_b:p_i<\epsilon\,,p_h\geq p_i\ \forall h\,,p_h>p_i\ \forall h<i\right\}\,
$$
for $1 \leq i \leq b$, that is, distributions in $\widehat{\mathcal{P}}_b^{i}$ have a minimum component in the $i$-th coordinate, which is smaller than $\epsilon$, and strictly smaller than any of the preceding components (unless of course $i=1$). Correspondingly, define a set of balanced distributions as
$$
 \widehat{\mathcal{P}}_b^{0} = \left\{p\in \mathcal{P}_b:p_i\geq \epsilon\ \forall i\right\}\,.
$$
The symmetry argument mentioned before also applies in this case and we can continue in analogy replacing the $m_{i,h}$ of \eqref{eq:ReducedQuadratic} with the following quantities
\begin{equation}
\begin{aligned}
\widehat{M}_1 & =\sup_{p,q \in \widehat{\mathcal{P}}_b^{0}} \Psi(p,q) &\qquad &
\widehat{M}_2 & =\sup_{p \in \widehat{\mathcal{P}}_b^{0},q\in \widehat{\mathcal{P}}_b^{i}} \Psi(p,q)\\
\widehat{M}_3 & =\sup_{p,q \in \widehat{\mathcal{P}}_b^{i}} \Psi(p,q) & \qquad &
\widehat{M}_4 & =\sup_{p \in \widehat{\mathcal{P}}_b^{i},q\in \widehat{\mathcal{P}}_b^{h}} \Psi(p,q)
\end{aligned}
\label{eq:MinMis}
\end{equation}
where again $1\leq i,h\leq b$ with $i\neq h$.

Applying the above scheme with the symmetric partitions we just defined, we can now rewrite the upper bound of equation \eqref{eq:ReducedQuadratic} in the form
\begin{align}
\sum_{p,q}&\lambda_p\lambda_q \Psi(p,q) \nonumber \\ & \leq \eta_0^2 M_1 + \eta_0\sum_{i>0}\eta_i M_2+\sum_{i>0}\eta_i^2 M_3+2\sum_{0<i<h}\eta_i\eta_h M_4\,.
\label{eq:QuadracBoundedbyMs}
\end{align}
Call $M$ be the maximum value achieved by the right hand side of \eqref{eq:QuadracBoundedbyMs} over all possible probability distributions $\eta=\eta_0,\eta_1,\ldots,\eta_b$ (which will of course depend on whether we use the $\widehat{M}_i$'s or $\widecheck{M}_i$'s values in place of the $M_i$'s). The optimization of \eqref{eq:QuadracBoundedbyMs}, once known the $M_i$'s values, is easy using the standard lagrange multipliers method (or see Lemma 2 of \cite{DellaFioreCostaDalai}).
Then we can then replace $\Psi_{\max}$ in \eqref{eq:RboundedbyPsi} with $M$ to derive the bound
\begin{equation*}
R_{(b,k)} \leq \frac{1}{\frac{2}{M \log \frac{b-j}{k-j-1}}+\frac{1}{\log\left(\frac{b}{j-1}\right)}}\,.
\label{eq:RboundedbyPsiMax}
\end{equation*}
We will describe in the next Section our procedure to determine, or upper bound the values $\widehat{M}_i$, $\widecheck{M}_i$ and the corresponding $M$. Here we only state the obtained results.

Using the partition based on the maximum value $\{\widecheck{\mathcal{P}}_b^{i}\}_{i=0,\ldots,b}$ we obtain the following theorem.
\begin{teo}
We have
\begin{align*}
	R_{(7, 7)} \leq& 0.0408975, \: R_{(8,8)} \leq 0.0188887, \: R_{(9,8)} \leq 0.0561537,
	\\
	 &R_{(10,9)} \leq 0.0277279, \: R_{(11,10)} \leq 0.0132033\,.
\end{align*}
\end{teo}
Using the partition based on the minimum value $\{\widehat{\mathcal{P}}_b^{i}\}_{i=0,\ldots,b}$ we obtain the following theorem.
\begin{teo}\label{BestBoundsBelow}
We have
\begin{align*}
	R_{(5,5)}\leq 0.1689&325, \qquad R_{(6,5)} \leq 0.3451130,
	\\
	R_{(6,6)} &\leq \frac{5}{59} \approx 0.0847458\,.
\end{align*}
\end{teo}

Based on the results in \cite{DalaiVenkatJaikumar2}, on its generalization given in equation \eqref{eq:dalaivenkatbound} and on Theorem \ref{BestBoundsBelow} when $(b,k)=(6,6)$, we are led to formulate the following conjecture.
\begin{con}\label{conj1}
For $b \geq k>3$, 
$$
	R_{(b,k)} \leq \min_{2 \leq j \leq k-2} \left(\frac{1}{\log\frac{b}{j-1}} + \frac{b^{j+1}}{b^{\underline{j+1}} \log \frac{b-j}{k-j-1}} \right)^{-1}\,.
$$
\end{con}
Note that the conjectured expression can be seen as a modification of the K\"orner-Marton bound in \eqref{eq:kornermarton} which takes into account the effects of prefix-based partitions.

\section{Computation of $M$}\label{CaseReduction}

Thanks to a straightforward generalization of some lemmas defined and proved in \cite{DellaFioreCostaDalai}, we have determined and inspected using Mathematica all the possible maximum points (see the Appendices in \cite{DellaFioreCostaDalai}) in which each $\widecheck{M}_i$ (or $\widehat{M}_i$) can be attained, obtaining the following propositions.
\begin{prop}\label{Mabove}
For $j=k-2$, we have that


\begin{table}[ht!]
\def\arraystretch{1.}
\centering
\begin{tabularx}{\linewidth}{c@{\extracolsep{\fill}}c@{\extracolsep{\fill}}c@{\extracolsep{\fill}}c@{\extracolsep{\fill}}c@{\extracolsep{\fill}}c@{\extracolsep{\fill}}l}
$(b,k)$ & $\epsilon$ & $\widecheck{M}_1$ & $\widecheck{M}_2$ & $\widecheck{M}_3$ & $\widecheck{M}_4$\\
\hline
$(7,7)$ & $9/100$ & 0.085679 & 0.092593 & 0.000006 & 0.000107\\
$(8,8)$ & $3/25$ & 0.038453 & 0.042840 & 0.000002 & 0.000022\\
$(9,8)$ & $1/10$ & 0.075870 & 0.076905 & 0.000001 & 0.000015\\
$(10,9)$ & $1/15$ & 0.036289 & 0.037935 & $3.4 \cdot 10^{-9}$ & $8.5 \cdot 10^{-8}$\\
$(11,10)$ & $1/11$ & 0.016928 & 0.018144 & $1.4 \cdot 10^{-9}$ & $2.7 \cdot 10^{-8}$\\
\hline
\vspace{-2.5mm}
\end{tabularx}
\def\arraystretch{1.35}
\begin{tabularx}{\linewidth}{l}
$\widecheck{M}_1$ attained at  $(\frac{1}{b}, \ldots, \frac{1}{b}; \frac{1}{b}, \ldots, \frac{1}{b})$\\
$\widecheck{M}_2$ attained at $(1, 0, \ldots, 0; 0, \frac{1}{b-1}, \ldots, \frac{1}{b-1})$\\
$\widecheck{M}_3$ attained at $(1-\epsilon, \frac{\epsilon}{b-1}, \ldots, \frac{\epsilon}{b-1}; 1-\epsilon, \frac{\epsilon}{b-1}, \ldots, \frac{\epsilon}{b-1})$\\
$\widecheck{M}_4$ attained at $(1-\epsilon, \frac{\epsilon}{b-2}, \ldots, \frac{\epsilon}{b-2}, 0; 0, \frac{\epsilon}{b-2}, \ldots, \frac{\epsilon}{b-2},  1-\epsilon)$\\
\hline
\end{tabularx}
\end{table}
\end{prop}

\newpage

\begin{prop}\label{Mbelow}
For $j=3$, $(b,k)=(5,5)$ and $\epsilon = \frac{1}{44}(4+\sqrt{5})$ we have that
%
\begin{table}[ht!]
\def\arraystretch{1.5}
\begin{tabularx}{\linewidth}{cl@{\extracolsep{\fill}}l}
$\widehat{M}_i$ & Attained at point $(p;q)$ & Values $\approx$ \\
\hline
$\widehat{M}_1$ & $(\epsilon,\frac{1-\epsilon}{b-1}, \ldots, \frac{1-\epsilon}{b-1}; \gamma, \delta, \ldots, \delta), \delta \approx 0.185275$ & 0.384033\\
\hline
$\widehat{M}_2$ & $(0, \frac{1}{b-1}, \ldots, \frac{1}{b-1}; \gamma, \delta, \ldots, \delta), \delta = \epsilon$ & 0.389226\\
\hline
$\widehat{M}_3$ & $(\epsilon, \frac{1-\epsilon}{b-2}, \ldots, \frac{1-\epsilon}{b-2}, 0; \epsilon, \alpha, \ldots, \alpha, \beta), \beta \approx 0.4542$ & 0.374759\\
\hline
$\widehat{M}_4$ & $(0, \frac{1}{b-1}, \ldots, \frac{1}{b-1}; \gamma, \delta, \ldots, \delta), \delta = \epsilon$ & 0.389226\\
\hline
\end{tabularx}
\end{table}

For $j=3$, $(b,k)=(6,5)$ and $\epsilon = \frac{1}{10}$ we have that
\begin{table}[ht!]
\def\arraystretch{1.5}
\begin{tabularx}{\linewidth}{cl@{\extracolsep{\fill}}l}
$\widehat{M}_i$ & Attained at point $(p;q)$ & Values $\approx$ \\
\hline
$\widehat{M}_1$& $(\epsilon,\frac{1-\epsilon}{b-1}, \ldots, \frac{1-\epsilon}{b-1}; \gamma, \delta, \ldots, \delta), \delta \approx 0.153159$ & 0.555625\\
\hline
$\widehat{M}_2$ & $(0, \frac{1}{b-1}, \ldots, \frac{1}{b-1}; \gamma, \delta, \ldots, \delta), \delta \approx 0.130217$ & 0.558467\\
\hline
$\widehat{M}_3$ & $(\epsilon, \frac{1-\epsilon}{b-2}, \ldots, \frac{1-\epsilon}{b-2}, 0; \epsilon, \alpha, \ldots, \alpha, \beta), \beta \approx 0.37693$  & 0.535106\\
\hline
$\widehat{M}_4$ & $(0, \frac{1}{b-1}, \ldots, \frac{1}{b-1}; \gamma, \delta, \ldots, \delta), \delta \approx 0.130217$ & 0.558467\\
\hline
\end{tabularx}
\end{table}

For $j=4$, $(b,k)=(6,6)$ and $\epsilon = \frac{1}{20}$ we have that
\begin{table}[ht!]
\def\arraystretch{1.5}
\begin{tabular*}{\linewidth}{cl@{\extracolsep{\fill}}l}
$\widehat{M}_i$ & Attained at point $(p;q)$ & Values $\approx$ \\
\hline
$\widehat{M}_1$& $(\frac{1}{b}, \ldots, \frac{1}{b}; \frac{1}{b}, \ldots, \frac{1}{b})$ & 0.185185\\
\hline
$\widehat{M}_2$ & $(\epsilon, \frac{1-\epsilon}{b-1}, \ldots, \frac{1-\epsilon}{b-1}; \gamma, \delta, \ldots, \delta), \delta \approx 0.147757$ & 0.178857\\
\hline
$\widehat{M}_3$ & $(\epsilon, 0, \frac{1-\epsilon}{b-2}, \ldots, \frac{1-\epsilon}{b-2}; 0, 1, 0, \ldots, 0)$  & 0.140664\\
\hline
$\widehat{M}_4$ & $(1, 0, \ldots, 0; 0, \frac{1}{b-1}, \ldots, \frac{1}{b-1})$ & $0.192000$\\
\hline
\end{tabular*}
\end{table}
\\
The values reported for $\widehat{M}_3$ are not approximate values of the exact values of $\widehat{M}_3$ but, instead, they are upper bounds.
\end{prop}

\begin{rem}
We point out that the value $\widehat{M}_1$ for $(b,k)=(6,6)$ is only attained for uniform distributions.
\label{rem:uniquemax}
\end{rem}

As a consequence of Propositions \ref{Mabove}, \ref{Mbelow} and equation \eqref{eq:QuadracBoundedbyMs} we are able to evaluate the values of $M$ for both the partitions $\{\widecheck{P}_b^i\}_{i=0,\ldots,b}$ and $\{\widehat{P}_b^i\}_{i=0,\ldots,b}$. Then we state the following theorem
\begin{teo}\label{theorem:Msvalues}
	Using the partition $\{\widecheck{P}_b^i\}_{i=0,\ldots,b}$ we get
	\begin{itemize}
		\item for $(b,k) = (7,7)$ we have that $M \approx 0.0861594$;
		\item for $(b,k) = (8,8)$ we have that $M \approx 0.0388599$;
		\item for $(b,k) = (9,8)$ we have that $M \approx 0.0758830$.
		\item for $(b,k) = (10,9)$ we have that $M \approx 0.0363565$.
		\item for $(b,k) = (11,10)$ we have that $M \approx 0.0170049$.
	\end{itemize}
	Using the partition $\{\widehat{P}_b^i\}_{i=0,\ldots,b}$ we get
	\begin{itemize}
		\item for $(b,k) = (5,5)$ we have that $M \approx 0.3873676$;
		\item for $(b,k) = (6,5)$ we have that $M \approx 0.5567010$;
		\item for $(b,k) = (6,6)$ we have that $M = \frac{5}{27} \approx 0.185185$.
	\end{itemize}
\end{teo}


For the values of $(b,k)$ reported in Table \ref{tab:bkbounds} except the cases in which $k=5$, $b=k=6,7,8$ and $(b,k) = (9,8)$, $(10, 9)$, $(11, 10)$, it is interesting to note that the bounds in bold (the generalized bounds \cite{costaDalai} or \cite{DalaiVenkatJaikumar}) are achieved for uniform distributions. This means that, for these particular cases, any new upper bounds that can be found on the quadratic form in equation \eqref{simmetrizzata2} cannot further improve those bounds. However, for such globally balanced codes, one can use a different argument based on the minimum distance of the code to get even stronger upper bounds. A proof that $R_{(6,6)} < 5/59$, based on the Aaltonen bound \cite{Aaltonen}, can be found in \cite{DellaFioreCostaDalai}.

\end{document}